\documentclass[aps,prl ,showpacs,twocolumn,groupedaddress,superscriptaddress,nobibnotes]{revtex4-1}

\usepackage{blindtext}

\usepackage[colorlinks,
          linkcolor=blue,
            citecolor=blue,
            urlcolor=blue
           ]{hyperref}
\usepackage{amsmath,amsthm,amssymb}
\usepackage{array}
\usepackage{subfigure}
\usepackage{graphicx}
\usepackage{fancyhdr}
\usepackage{color}
\usepackage{extarrows}
\usepackage{enumerate}
\usepackage{epstopdf}
\usepackage{bm}
\usepackage{natbib}
\usepackage{comment}
\usepackage{float}
\usepackage{ulem}

\begin{document}

\title{Giant Linear Dichroism Controlled by Magnetic Field in FePS$_3$}

\author{Xu-Guang~Zhou}
\thanks{These authors contributed equally to this work.}
\affiliation{Institute for Solid State Physics, University of Tokyo, Kashiwa, Chiba 277-8581, Japan}

\author{Zhuo~Yang}
\thanks{These authors contributed equally to this work.}
\affiliation{Institute for Solid State Physics, University of Tokyo, Kashiwa, Chiba 277-8581, Japan}

\author{Youjin~Lee}
\affiliation{Center for Quantum Materials, Seoul National University, Seoul 08826, Republic of Korea}
\affiliation{Department of Physics and Astronomy, Institute of Applied Physics, Seoul National University, Seoul 08826, Republic of Korea}

\author{Jaena Park}
\affiliation{Center for Quantum Materials, Seoul National University, Seoul 08826, Republic of Korea}
\affiliation{Department of Physics and Astronomy, Institute of Applied Physics, Seoul National University, Seoul 08826, Republic of Korea}

\author{Yoshimitsu~Kohama}
\affiliation{Institute for Solid State Physics, University of Tokyo, Kashiwa, Chiba 277-8581, Japan}

\author{Koichi~Kindo}
\affiliation{Institute for Solid State Physics, University of Tokyo, Kashiwa, Chiba 277-8581, Japan}

\author{Yasuhiro~H.~Matsuda}
\affiliation{Institute for Solid State Physics, University of Tokyo, Kashiwa, Chiba 277-8581, Japan}

\author{Je-Geun~Park}
\affiliation{Center for Quantum Materials, Seoul National University, Seoul 08826, Republic of Korea}
\affiliation{Department of Physics and Astronomy, Institute of Applied Physics, Seoul National University, Seoul 08826, Republic of Korea}

\author{Oleg~Janson}
\email{o.janson@ifw-dresden.de}
\affiliation{Institute for Theoretical Solid State Physics, Leibniz IFW Dresden, 01069 Dresden, Germany}

\author{Atsuhiko~Miyata}
\email{a-miyata@issp.u-tokyo.ac.jp}
\affiliation{Institute for Solid State Physics, University of Tokyo, Kashiwa, Chiba 277-8581, Japan}

\begin{abstract}
Magnetic-field control of fundamental optical properties is a crucial challenge in the engineering of multifunctional microdevices. 
Van der Waals (vdW) magnets retaining a magnetic order even in atomically thin layers, offer a promising platform for hosting exotic magneto-optical functionalities owing to their strong spin-charge coupling. 
Here, we demonstrate that a giant optical anisotropy can be controlled by magnetic fields in the vdW magnet FePS$_3$. The giant linear dichroism ($\sim$11\%), observed below $T_{\text{N}}\!\sim\!120$\,K, is nearly fully suppressed in a wide energy range from 1.6 to 2.0\,eV, following the collapse of the zigzag magnetic order above 40\,T. This remarkable phenomenon can be explained as a result of symmetry changes due to the spin order, enabling minority electrons of Fe$^{2+}$ to hop in a honeycomb lattice. The modification of spin-order symmetry by external fields provides a novel route for controllable anisotropic optical micro-devices.
\end{abstract}
\maketitle

Optical anisotropy constitutes a fundamental and crucial concept for linear and
nonlinear optical components such as polarizers, wave plates, and
phase-matching elements.  It is a powerful tool not only for providing direct
information on symmetry breaking but also as a key link in describing the
quantum state of photons within the quantum mechanics
framework~\cite{sakurai1995modern}.  To generate linearly polarized light,
unpolarized light typically passes through a crystal with structural or
electronic anisotropy.  These anisotropies are achieved by using materials
such as liquid crystals, polymers, and some birefringent
crystals~\cite{norden1977linear,niu2018giant}.  While conventional materials
providing linear polarizations have been successfully commercialized for many 
applications, their linear dichroism (LD) remains constant and uncontrollable. In particular, their intrinsic three-dimensional structure prevents 
its incorporation into micro-devices.  Recent studies of two-dimensional materials
with large LD effects provide a completely new path for the realization of 
micro-devices, such
as the black phosphorus (BP)~\cite{mao2016optical} and rhenium disulfide
(ReS$_2$)~\cite{wang2019plane}.  Despite the excellent potential for
application, reports on microdevices capable of controlling optical anisotropy
remain scarce~\cite{zhang2022cavity}.

In contrast to traditional three-dimensional magnetic materials, 
van der Waals (vdW) magnets, which have been intensively studied recently, 
have a huge potential for developing micro-controllable optical 
devices.  Like graphite, these materials can be exfoliated into a single
layer, exhibiting fascinating mesoscopic properties, and maintaining bulk-like
magnetic and multiferroic behavior even in a single
monolayer~\cite{coak2019tuning, chittari2016electronic}.  This makes them easily stackable and allows
for fabrication of functional heterostructures for micro-controllable devices.
Remarkably, in vdW magnets, e.g., $M$P$X$$_3$ family, where $M$ is a $3d$
transition element and $X$ is S or Se~\cite{coak2019tuning}, strong couplings between magnetism and
electronic properties have been recently reported, leading to exotic optical
properties such as coherent many-body excitons in NiPS$_3$ and giant linear
dichroism in FePS$_3$~\cite{kang2020coherent,zhang2021observation}.  
In addition, external magnetic fields can play a crucial role
in modifying such strong couplings and give rise to the emergence of novel
magneto-optical properties in vdW magnets.

FePS$_3$ has a monoclinic structure with $C2/m$ symmetry. Below $T_{\text{N}}\sim\!120$\,K, it features an Ising-type zigzag order with magnetic chains running along the $a$ axis~\cite{kurosawa1983neutron,koitzsch2023intertwined, lee2016ising,peng2023ferromagnetism}. 
Previous studies have revealed a giant LD effect near the band-gap energy ($\sim1.4\text{-}1.5$~eV~\cite{koitzsch2023intertwined}) in this compound only below $T_\text{N}$~\cite{zhang2021observation,zhang2022cavity,koitzsch2023intertwined}, 
indicating that the magnetic order is crucial for the LD effect.
Interestingly, using micro-nano fabrication techniques, a cavity-coupled FePS$_3$ exhibits the giant LD reaching up to almost 100\%~\cite{zhang2022cavity}.
Furthermore, recent high-field magnetization experiments under an out-of-plane field 
revealed a 
sudden jump in magnetization at 40\,T~\cite{wildes2020high}. 
This is understood as an Ising-like spin-flip transition in FePS$_3$, where the
spin configurations change from a zigzag antiferromagnetic order to a
spin-polarized phase. The step-like magnetization curve implies that an
external field is a promising way to realize the switch-like control of the
giant LD effect.

\begin{figure*}[t]
    \begin{center}
    \includegraphics[width = 1.0\linewidth]{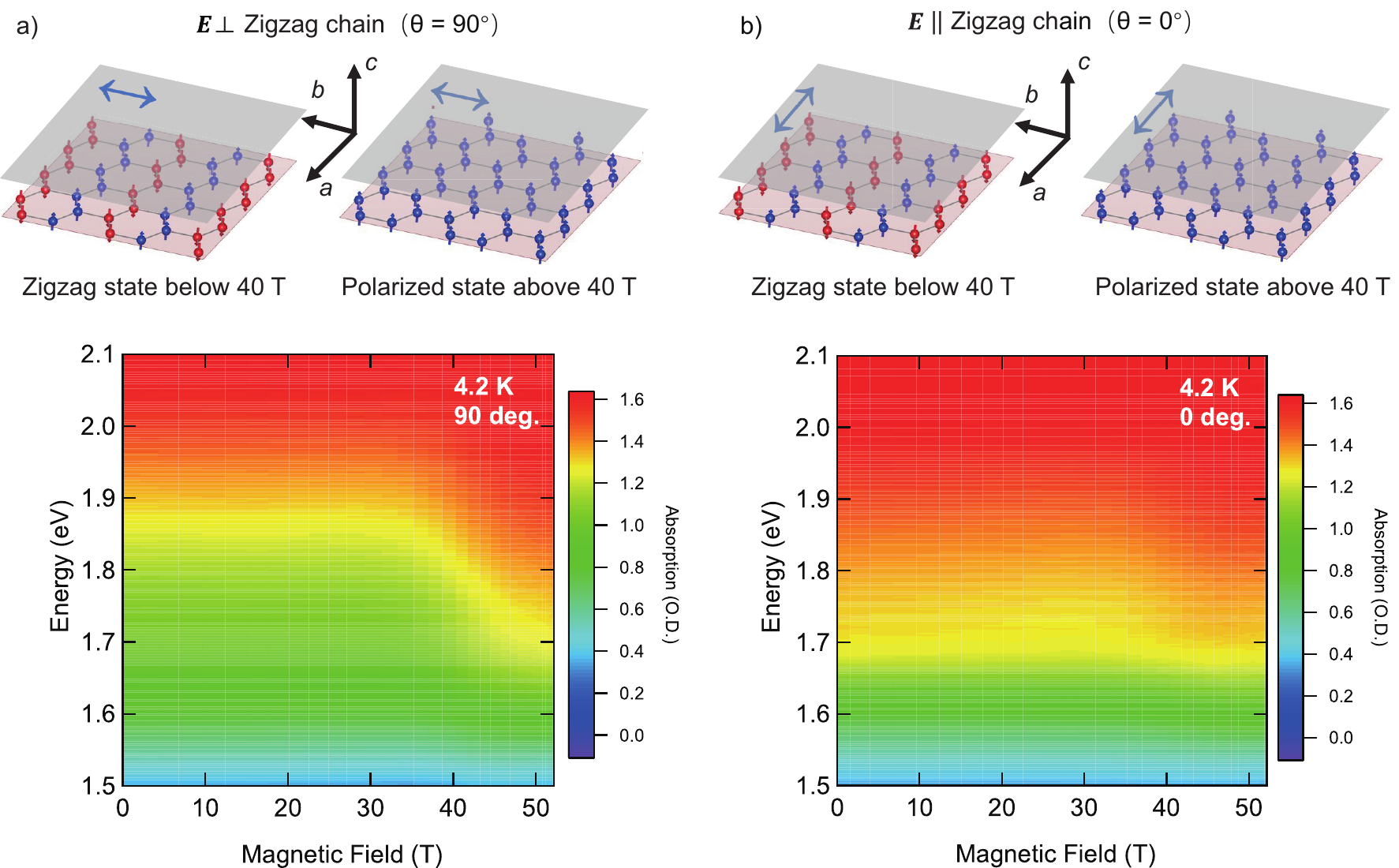}
     \caption{The field-dependent absorption spectra of FePS$_3$ in two
geometries, where zigzag chains are a) perpendicular and b) parallel to the
$\boldsymbol{E}$ polarization. The gray squares represent the neutral density
linear polarizer, and the directions of $\boldsymbol{E}$ is marked with the blue
arrows. The spin order of an exfoliated multi-layered sample is drawn by the
honeycomb lattice bellow the polarizer. The spectra are obtained at 4.2\ K in
helium gas environment and the external field is parallel to the $c$-axis.}
     \label{2Dimage}
    \end{center}
    \end{figure*}

In this study, we measured the magnetic field dependence of the LD effect in FePS$_3$ 
with $\boldsymbol{B}$ of up to 52 T ($\boldsymbol{B} \parallel$ $c$ axis). 
The results demonstrate a sudden jump in LD at 40 T, 
coinciding with the breakdown of zigzag spin order symmetry. 
Our findings support the notion that spin order symmetry is the primary factor 
influencing the LD effect. Furthermore, following the substantial jump in LD under 
an external field, we propose that a vdW antiferromagnet, FePS$_3$, can serve 
as a microdevice for the switch-like control of an optical anisotropy under a magnetic field. Given the 
precision and rapid regulation capabilities of external fields compared to temperature, 
such optical devices are anticipated to play an active role in quantum optics and optical computing. Note that cm-sized pulsed magnets are nowadays widely available.

FePS$_3$ flakes are mechanically exfoliated from a bulk crystal using Scotch tape and transferred 
onto the SiO$_2$ substrate. The size of SiO$_2$ substrate is 2 mm in diameter and 0.3 mm in thickness.
The thickness of a FePS$_3$ sample is estimated to be 2-3~$\mu$m by the interference effect~\cite{yang2021}.
Magneto-absorption measurements are performed with a pulse magnet for magnetic fields 
up to 52~T. The pulse duration of the magnetic field is about 36~ms. The field direction and the propagation direction of light are along the $c$-axis of 
the FePS$_3$ sample (the Faraday configuration). The sample was mounted in
a helium gas-filled tube which was placed in a liquid helium cryostat to achieve a low temperature of 4.2~K. 
A broadband halogen lamp is employed as the light source. The light emitted from the lamp is coupled to an 
800 $\mu$m core diameter multimode fiber, used to illuminate the sample. The linear polarizer is placed 
between the excitation fiber and the sample, and the distance between the polarizer and the sample is 
less than 1 mm. The transmitted light is coupled in an 800 $\mu$m diameter multimode fiber and guided 
to a spectrometer equipped with a CCD camera. The typical exposure time of 0.5 ms is much shorter than the field duration, ensuring
that the transmission spectra were acquired at essentially constant magnetic field values.

Figure~\ref{2Dimage} illustrates the field-dependent absorption spectra of FePS$_3$ in two 
distinct geometries, both acquired at 4.2~K. The absorption $A$ is obtained by $-\text{log}_{10}\frac{I}{I_0}$, 
where $I$ is a transmitted intensity and $I_0$ an incident intensity.
In the first geometry, as shown in Fig.~\ref{2Dimage}(a), 
the zigzag chain ($a$ axis) is perpendicular to the electric-field polarization of light $\boldsymbol{E}$, achieved using a neutral density linear polarizer. 
On the other hand, the second geometry, as shown in Fig.\ref{2Dimage}(b), presents the parallel condition. 
The angle between $\boldsymbol{E}$ and the zigzag chain ($a$-axis) is denoted as $\theta$. When $\theta$ equals 90$^{\circ}$, 
the absorption spectra remain unchanged below an external field of 40 T. 
However, beyond this threshold, remarkably,  the optical absorption in FePS$_3$ experiences a sudden enhancement around 1.7~eV. 
This behavior contrasts with the $\theta=0^{\circ}$ case, 
as shown in Fig~\ref{2Dimage}(b), where the absorption spectra are almost field-independent up to 52~T.

\begin{figure}[t]
    \begin{center}
    \includegraphics[width = 9.0cm]{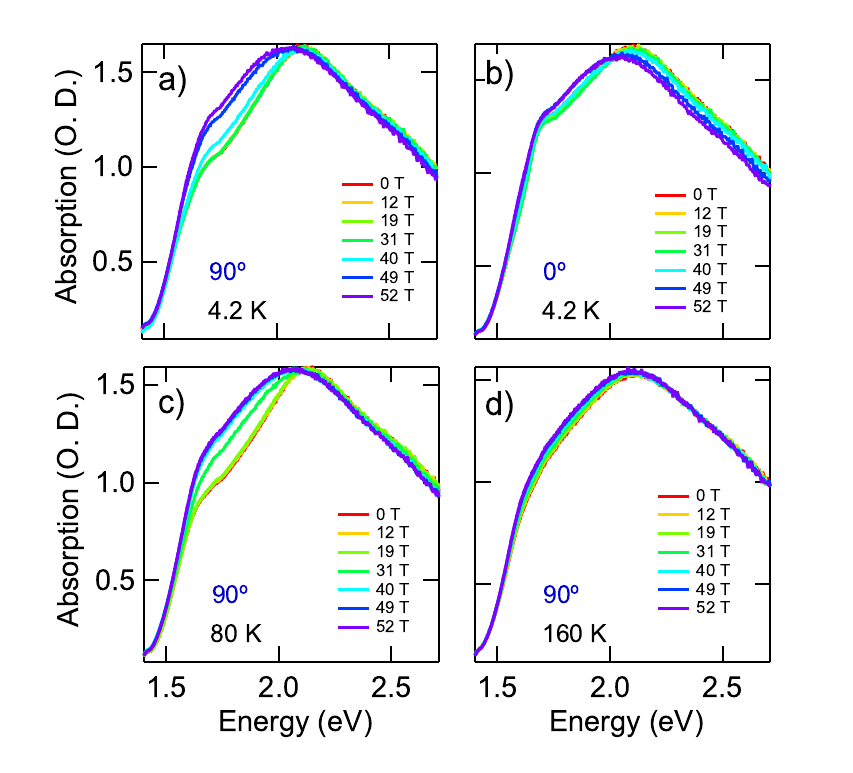}
     \caption{Absorption spectra for different polarization angles and temperatures in magnetic field;
     a) 90$^{\circ}$ and 4.2~K,  b) 0$^{\circ}$ and 4.2~K, c) 90$^{\circ}$ and 80~K, and c) 90$^{\circ}$ and 160~K.
     The change of absorption is only observed for $\theta$ is 90$^{\circ}$ below $T_{\text{N}}$.}
     \label{Spectra}
    \end{center}
    \end{figure}

Differences between the two polarization angles become evident in 1D plots, where absorption is plotted as a function of energy by applying a magnetic field.
In Fig.~\ref{Spectra}(a), i.e., $\theta$ = 90$^{\circ}$, the optical
absorptions between 1.6~eV and 2.0~eV are strongly suppressed without a
magnetic field. 
This suppression remains almost intact up to the full magnetic polarization at 40~T, at which
the absorption spectrum experiences a rapid enhancement before reaching saturation.
Outside this energy range, no clear field dependence of the absorption
spectrum is observed.  In contrast, for $\theta$ = 0$^{\circ}$, the optical
absorption between 1.6 and 2.0~eV is not suppressed at 0 T 
(i.e., the observation of the giant LD effect). For $\theta$ = 0$^{\circ}$,
almost no magnetic field dependence is observed (Fig.~\ref{Spectra}, b).

 \begin{figure}[b]
    \begin{center}
    \includegraphics[width = 9.0cm]{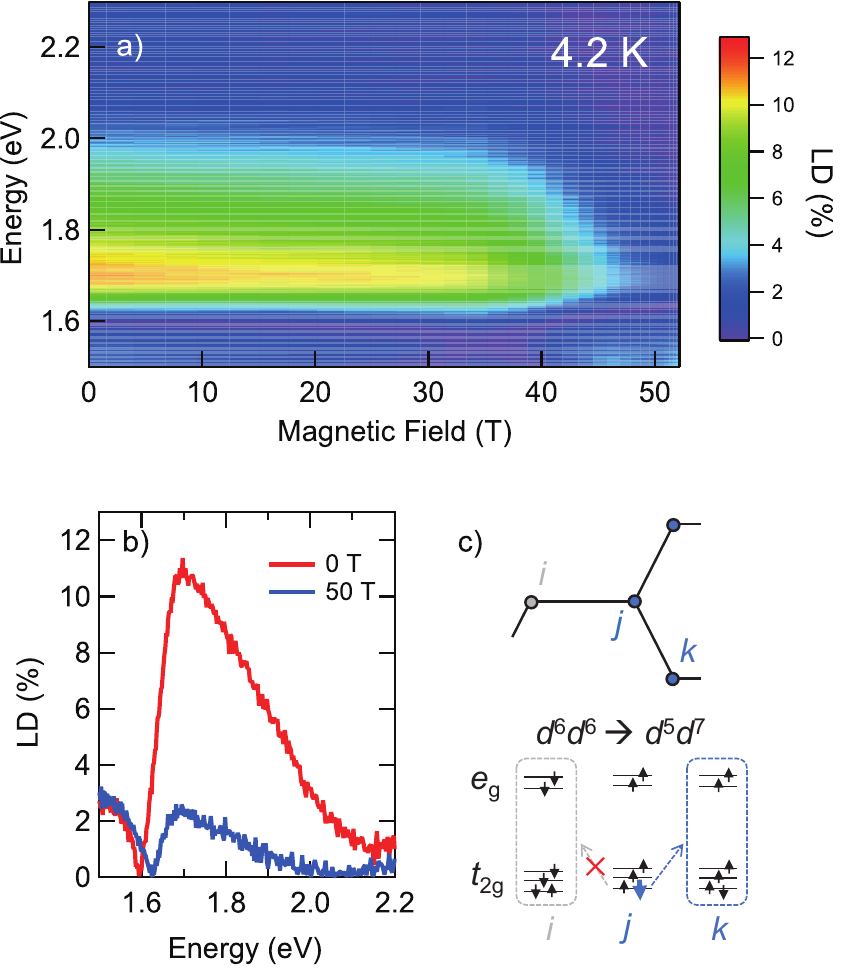}
     \caption{a) Field dependent color mapped LD spectra up to 52 T. b) The comparison of LD spectra between 0 and 50~T. c) The schematic diagram of spin hopping in the zigzag state, that can explain the origin of the strong dichroism observed in the absent of magnetic field.  
     The minority spins (electrons) are marked with big blue arrow.}
     \label{Compr.LD}
    \end{center}
    \end{figure}

In order to better understand the impact of spin order on the field-induced change of the absorption spectrum, 
we investigate the magneto-absorption spectra of FePS$_3$ at different temperatures below and above $T_\text{N}$ $\sim$ 120 K. 
In Fig.~\ref{Spectra}(c) and (d), the magneto-absorption spectra 
at 80~K and 160~K are presented for $\theta$ = 90$^{\circ}$ ($\boldsymbol{E}$ perpendicular to the zigzag chain).
It is noteworthy that the field induced-change in the absorption spectra at
80~K is similar to that observed at 4.2~K, although there is a subtle
reduction of a critical magnetic field to $\sim$30 T, where the absorption is drastically enhanced.  Above $T_\text{N}$, the effect of the
magnetic field on the absorption spectra vanishes. In other words, even at 0~T, the absorption spectrum remains unsuppressed, and its 
shape remains almost intact up to 52~T in the magnetically ordered state.
For comparison, detailed results of $\theta$ = 0$^{\circ}$ are shown in Figure S1.

We calculated the LD using $\frac{|A_{90^{\circ}} -
A_{0^{\circ}}|}{A_{90^{\circ}} + A_{0^{\circ}}} \times 100\%$ to directly 
see the change of LD in FePS$_3$ controlled by magnetic fields~\cite{zhang2021observation}. In Fig.~\ref{Compr.LD}(a), 
we show the color mapping of the LD obtained at 4.2~K. At 0 T, 
we observe the giant LD between 1.6 and 2.0~eV as previously reported. 
This LD shows almost unchanged below 40~T and suddenly drops at 40~T and eventually reaches
zero around 52~T.  We compare the LD at 0 and 52~T in Fig.~\ref{Compr.LD}(b). 
The maximum of LD (almost 11\%) is observed at 1.75~eV, 
which can be controlled by a magnetic field.

To understand the drastic reduction of the dichroic signal, we first discuss
the origin of the strong dichroism observed in the zero field.  As shown in
Fig.~\ref{Compr.LD}(c), in the high-spin $d^6$ configuration of Fe$^{2+}$, the
one spin channel is fully filled, while the other is occupied by a single
electron.  In the N\'eel antiferromagnetic state, the Pauli principle makes
this electron immobile: the corresponding spin sector is fully occupied in each
of the three neighboring sites.  Hence, the nearest-neighbor hopping is
facilitated by majority electrons, i.e. the up spin electrons at sites $j$ and $k$, or 
the down spin electrons at site $i$. But in the zigzag antiferromagnetic ground
state of FePS$_3$ the situation is very different.  Here, each Fe atom has two
neighbors with a parallel magnetic moment (along the zigzag chains) and one
neighbor whose moment is antiparallel. Only the majority electrons can hop to the
latter; by the same argument, the in-chain hopping processes are restricted to
minority electrons.  While both types of processes are possible, the Hund
exchange gives rise to disparity: a minority electron leaves the energetically
favorable high-spin state behind, while a majority electron creates the
low-spin state with high energy. Thus, hopping along the chains dominates the
low-energy sector of optical excitations (dipole transitions) in the zigzag
state~\cite{koitzsch2023intertwined}.

Using the same reasoning, we can expect that in the fully polarized
ferromagnetic state, the low-lying optical excitations are facilitated by
minority states, and dichroism is absent. However, the actual electronic
structure of FePS$_3$ is more intricate. First, relevant hopping processes are
not restricted to nearest neighbors. Hopping between the second nearest neighbor via 
coupling across the honeycomb voids is
also present, as evidenced by the sizable magnetic exchange $J_3$ associated with
this path. Second, for the Fe$^{2+}$ orbital degrees of freedom are generally
relevant; their fingerprint is the giant magnetic anisotropy of 22~meV/Fe measured by
photoemission electron microscopy~\cite{lee2023giant}.  Finally,
two-dimensional superlattices of FePS$_3$ show sizable magnetoelastic
coupling~\cite{liu21}; structural changes at high magnetic fields can not be
excluded. Hence, direct calculations of the optical conductivity in the fully
polarized state are desirable for a microscopic picture.

To this end, we perform first-principles density-functional theory (DFT)
calculations using the generalized gradient approximation (GGA)~\cite{PBE96} as
implemented in the full-potential code FPLO version 22~\cite{FPLO}. For the
structural input, we use the experimental crystal structure~\cite{brec80} and
construct two different supercells that allow for N\'eel-orbital~\cite{suppl} and zigzag
magnetic order. For GGA+$U$ calculations, we used the Coulomb repulsion of 2.5~eV, the Hund exchange of 1 eV, and the fully localized limit as the double
counting correction.  This choice of interaction parameters 
follows~\cite{koitzsch2023intertwined}; other DFT+$U$ studies~\cite{zheng2019, olsen2021, nitschke2023, amirabbasi2023} employed similar parameters.
For both cells, the Brillouin zones were uniformly
sampled with approximately 2000 $k$-points.  Optical conductivities of zigzag
and ferromagnetic states were computed using the built-in optics module of
FPLO. In all calculations, dipole forbidden $d$-$d$ transitions are absent,
because the DFT's one-electron framework cannot describe many-body physics
underlying these processes.

\begin{figure}[tb]
  \includegraphics[width=8.6cm]{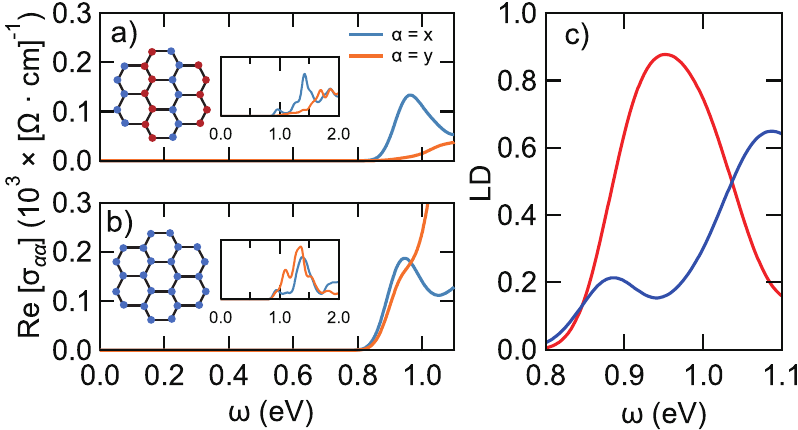}
  \caption{\label{fig:calc}Real part of optical conductivity $\sigma$ in the
zigzag antiferromagnetic (a) and ferromagnetic (b) states. The $x$ axis ($a$ axis) is
parallel to the zigzag chains and the $y$ axis ($b$ axis) is perpendicular to both $x$ and
the normal to the magnetic plane. A Gaussian broadening of
0.05\,eV is applied.  Note that the absorption edge in DFT+$U$ calculations is controlled by the Coulomb repulsion $U$; for the chosen $U$ = 2.5\,eV it is at $\sim$0.8 eV which is significantly lower than in the experiment ($\sim$1.4 eV). Dichroic ratios for the zigzag antiferromagnetic (red curve) and the ferromagnetic states (blue curve) are compared in (c). Here, the dichroic ratios (LD) are obtained as
$\frac{|Re[\sigma_{xx}] - Re[\sigma_{yy}]|}{Re[\sigma_{xx}] + Re[\sigma_{yy}]}$.
}
\end{figure}

In Fig.~\ref{fig:calc}, we compare the real part of optical conductivity
$\operatorname{Re}\left[\sigma_{\alpha\alpha}(\omega)\right]$ for the
zigzag state and the ferromagnetic state.  In the former case, the absorption
edge for the zigzag chain direction ($\alpha = x = a$-axis) is shifted to lower
frequencies compared to the transverse direction ($\alpha = y = b$-axis). This shift gives
rise to a dichroic signal, which reaches its maximal value close to the edge, but
remains sizable up to 1.5\,eV. The optical conductivity of the ferromagnetic
state is very different: $\operatorname{Re}\left[\sigma_{yy}\right]$ is shifted
to lower frequencies such that the absorption edge occurs at the same frequency
for both directions.  At the same time,
$\operatorname{Re}\left[\sigma_{xx}\right]$ does not change significantly in
near the absorption edge. Both features agree
with the experiment (Fig.~\ref{2Dimage}). The large difference in the absolute
values of dichroism can be attributed to domain structure in the samples.

For higher frequencies, the ferromagnetic state exhibits at odds with 
the experiment. While we cannot ultimately determine
the origin of this discrepancy, a structural transition accompanying the
magnetic polarization is a plausible scenario. Indeed, a strong magnetoelastic coupling 
is suggested from the previous high-field measurements concluded on the first order of 
the magnetic transition at 40~T~\cite{wildes2020high}. 
For such effects to be accounted for in calculations, the information on the
high-field crystal structure -- absent so far due to technical difficulties -- is needed~\cite{ikeda2022}.  
Another possible source of discrepancies is the neglect of the spin-orbit coupling: 
the chosen DFT code (FPLO) does not support calculations of optical spectra in the
full-relativistic mode. Both of which, experimental and theoretical, are the subject of 
future studies.

In summary, we experimentally demonstrated that a magnetic field can switch the
giant LD in the vdW antiferromagnet FePS$_3$.  Below $T_\text{N}$$\sim$120~K,
the LD of $\sim$11\% is suppressed to almost zero in a wide energy range from
1.6 to 2.0~eV by a magnetic field above 40~T. The vanishing LD is associated
with the change in the magnetic structure: the fully saturated state allows for
a more isotropic hopping of Fe$^{2+}$ minority electrons within the magnetic
planes. This scenario is supported by DFT calculations that further hint at a
first-order structural transition accompanying magnetic saturation.
Recently LD has been widely used as a powerful technique to detect symmetry changes in atomically thin layers. Our study provides significant insights into the coupling between spin-order symmetry and interatomic optical transitions.
The modification of spin-order symmetry by external fields is a new route to engineer multifunctional micro-devices of vdW materials.

Since FePS$_3$ maintains its magnetic properties even in a single monolayer, 
the remaining challenge is the reduction of the transition field in FePS$_3$. To this
end, choosing an appropriate substrate that induces a moderate tensile strain
in atomically thin FePS$_3$ is a promising route, as this alternation
particularly weakens the long-range antiferromagnetic cross-hexagon exchange
$J_3$~\cite{olsen2021}, which in turn governs the critical field.

\begin{acknowledgments}
We acknowledge Chaebin Kim for his generous assistance to the project.  X.-G.Z
was supported by a JSPS fellowship and funded by JSPS KAKENHI,
Grant-in-Aid for JSPS Fellows No. 22F22332.  A.M. was funded by JSPS KAKENHI,
Fund for the Promotion of Joint International Research (Home-Returning
Researcher Development Research) No. 22K21359.  O.J. was supported by the
Deutsche Forschungsgemeinschaft (DFG, German Research Foundation) Projects No.
247310070 and 465000489, and thanks U. Nitzsche for technical assistance.  This
work is also partially supported by JSPS KAKENHI, Grant-in-Aid for
Transformative Research Areas (A) No.23H04859, Grant-in-Aid for Scientific
Research (B) No. 18H01163, and the grant from Research Foundation for Opto-Science and 
Technology (Japan). The work at SNU is funded by the Leading Researcher
Program of the National Research Foundation of Korea (Grant No.
2020R1A3B2079375).
\end{acknowledgments}

\sloppy

%

\end{document}